# $a_0(980)$ photoproduction in the coupled channel model

Ł. Bibrzycki[1],[*] and R. Kamiński[2]

[1]*Institute of Computer Science, Pedagogical University of Cracow, Podchorążych 2, 30-084 Kraków, Poland*
[2]*H. Niewodniczański Institute of Nuclear Physics, Polish Academy of Sciences, Radzikowskiego 152, 31-342 Kraków, Poland*

**Abstract.** We calculated the coupled channel photoproduction amplitudes of the scalar isovector resononace $a_0(980)$, which include the $\pi\eta$, $K^+K^-$ and $K^0\overline{K^0}$ intermediate states. Based on them we obtained the mass distribution in the $\pi\eta$ channel at photon energy $E_\gamma$=7 GeV, corresponding to energy accissible in new JLab experiments CLAS12 and GlueX. We also analyzed the shape of mass distribution as a function of the phase of the background amplitude.

## 1 Introduction

As dominating decay channels of the $a_0(980)$ are $\pi\eta$ and $K\overline{K}$, the construction of the amplitudes of the resonant $\pi\eta$ photoproduction is inevitably the coupled channel problem. The model described in this paper assumes that the isovector $a_0(980)$ resonance observed in the $\pi\eta$ channel results from the $\pi\eta$ and $K\overline{K}$ interactions in the final state [1]. We describe the $a_0(980)$ photoproduction as a two stage process shown schematically in Fig.1. First, a pair of pseudoscalar mesons $\pi\eta$, $K^0\overline{K^0}$ or $K^+K^-$ is photoproduced, which we describe in terms of Born amplitudes. The final state interactions are described in terms of the coupled channel and unitary amplitudes. The general form of the $\pi\eta$ photoproduction amplitude, which takes into account the final state interactions is

$$\langle \lambda'|A_{\pi\eta}|\lambda_\gamma\lambda\rangle = \langle \lambda'|V_{\pi\eta}|\lambda_\gamma\lambda\rangle + 4\pi\sum_{\{mm'\}}\int_0^\infty \frac{k'^2 dk'}{(2\pi)^3} F(k') t_{\pi\eta;mm'} G_{mm'} \langle \lambda'|V_{mm'}|\lambda_\gamma\lambda\rangle \tag{1}$$

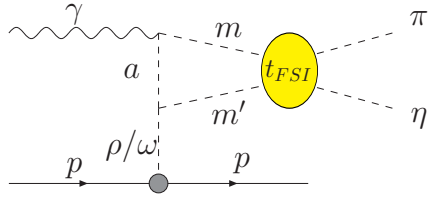

**Figure 1.** Diagram for the photoproduction process with final state interactions

---
[*]e-mail: lukasz.bibrzycki@ifj.edu.pl

where $V_{\pi\eta}$ ($V_{mm'}$) is the Born amplitude of the $\pi\eta$ ($mm'$) pair photoproduction, $t_{\pi\eta;mm'}$ is the coupled channel scattering amplitude, $\lambda$, $\lambda'$ and $\lambda_\gamma$ are respectively the helicities of the initial and final proton, and photon helicity. $G_{mm'}$ is the propagator of the intermediate $mm'$ pair defined as

$$G_{mm'} = \frac{-1}{M_{mm'} - M'_{mm'}(k') + i\epsilon} \quad (2)$$

and $F(k')$ is the form-factor needed to regularize the divergent mesonic loop of the diagram shown in Fig.1. In present calculations we limit ourselves to the on-shell part of the amplitude and are concerned with the $S$-wave, so projecting out the higher partial waves, we arrive at the following coupled channel $S$-wave $\pi\eta$ photoproduction amplitude:

$$A_{\pi\eta} = \left[1 + i r_{\pi\eta} t^{I=1}_{\pi\eta}\right] V_{\pi\eta} + r_{K\overline{K}} t^{I=1}_{\pi\eta;K\overline{K}} \frac{1}{\sqrt{2}} \left(V_{K^+K^-} + V_{K^0\overline{K^0}}\right), \quad (3)$$

where $r_{\pi\eta} = -k_{\pi\eta} M_{\pi\eta}/8\pi$, $r_{K\overline{K}} = -k_{K\overline{K}} M_{\pi\eta}/8\pi$, $M_{\pi\eta}$ is effective mass of the $\pi\eta$ system, $t^{I=1}_{\pi\eta}$ and $t^{I=1}_{K\overline{K};\pi\eta}$ are isovector elastic $\pi\eta$ scattering amplitude and $K\overline{K} \to \pi\eta$ transition amplitude respectively. The $k_{\pi\eta}$ and $k_{K\overline{K}}$ are CM momenta in the $\pi\eta$ and $K\overline{K}$ channels defined as

$$k_{\pi\eta} = \frac{\sqrt{[M^2_{\pi\eta} - (m_\eta + m_\pi)^2][M^2_{\pi\eta} - (m_\eta - m_\pi)^2]}}{2M_{\pi\eta}} \quad \text{and} \quad k_{K\overline{K}} = \sqrt{\frac{M^2_{\pi\eta}}{4} - m^2_K}. \quad (4)$$

Further we use numerical channel indices for the elements of the final state scattering amplitude and denote $t_{\pi\eta}$ as $T_{11}$ and $t_{\pi\eta;K\overline{K}}$ as $T_{12}$ respectively.

## 2 Born amplitudes

In Fig. 2 we show the general structure of Feynman diagrams representing Born amplitudes. In these diagrams $a$ corresponds to $\rho$ or $\omega$ in the $\pi\eta$ channel and $K$ or $K^*$ in the $K\overline{K}$ channel. $b$ corresponds to the $\rho$ and $\omega$ exchange. Inclusion of diagram $c$) is necessary for electromagnetic current conservation when charged kaon pair is produced in the intermediate state. The amplitudes corresponding to

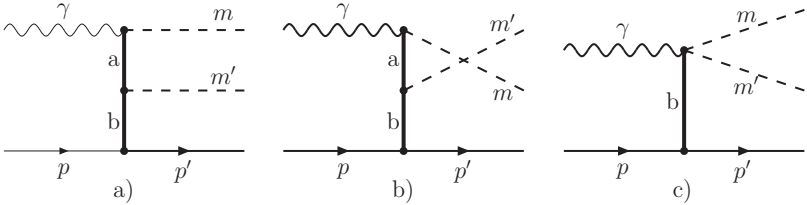

**Figure 2.** Diagrammatic representation of the Born amplitudes for $\pi\eta$, $K^0\overline{K^0}$ and $K^+K^-$ intermediate states.

diagrams of Fig.2 have a general form of:

$$V_{mm'} = \sum_{\{mm'\}} \overline{u}(p', s') J_{r,mm'} \cdot \varepsilon(q, \lambda^\gamma) u(p, s), \quad (5)$$

where the sum is over intermediate states $mm'$, $J_{r,mm'}$ is the hadronic current, $u(p, s)$ and $\overline{u}(p', s')$ - wave functions of the initial and final proton respectively and $\varepsilon$ - polarisation 4-vector of the incident photon. Details of the partial wave expansion of the Born amplitudes are discussed in [1, 2].

## 3 Final state interaction amplitudes

To obtain the final state rescattering amplitudes we solve coupled channel Lippman-Schwinger equations which in momentum space read [3–5]

$$\langle q|\hat{T}|k\rangle = \langle q|\hat{V}|k\rangle + \int \frac{d^3p}{(2\pi)^3} \langle q|\hat{V}|p\rangle \langle p|\hat{G}|p\rangle \langle p|\hat{T}|k\rangle, \tag{6}$$

where $\hat{V}$, $\hat{G}$ and $\hat{T}$ are 2×2 matrices in channel space. $\hat{G}$ denotes the diagonal propagator matrix, with the elements defined as

$$G_i(p) = \frac{1}{E - E_i(p) + i\epsilon}. \tag{7}$$

$\hat{V}$ is the interaction matrix which is assumed in a separable form

$$\langle q|V_{ij}|k\rangle = \lambda_{ij} g_i(q) g_j(k), \tag{8}$$

where $\lambda_{ij}$ is a 2×2 coupling constant matrix and the form factor is defined as

$$g_i(p) = \sqrt{\frac{4\pi}{m_i}} \frac{1}{p^2 + \beta_i^2}. \tag{9}$$

The on-shell scattering matrix elements $T_{ij}(k_i, k_j)$ can be expressed in terms of the $S$-matrix elements

$$S_{ij} = \delta_{ij} - \frac{i}{\pi} \sqrt{k_i \alpha_i k_j \alpha_j}\, T_{ij}(k_i, k_j) \tag{10}$$

where $i$ and $j$ enumerate channels, $k_i$, $k_j$ are center of mass momenta in respective channels and $\alpha_i$ are defined as:

$$\alpha_1 = \frac{E_\pi E_\eta}{E_\pi + E_\eta} \tag{11}$$

and

$$\alpha_2 = \frac{E_K}{2}. \tag{12}$$

Here $E_\pi$, $E_\eta$ and $E_K$ denote the $\pi$, $\eta$ and $K$ energies in the center of mass of respective channels. For the 2-channel case, the elements of the $S$-matrix can be related to the Jost function with the following expressions [4]:

$$S_{11} = \frac{D(-k_1, k_2)}{D(k_1, k_2)}, \tag{13}$$

$$S_{22} = \frac{D(k_1, -k_2)}{D(k_1, k_2)}, \tag{14}$$

$$S_{12}^2 = S_{11} S_{22} - \frac{D(-k_1, -k_2)}{D(k_1, k_2)}. \tag{15}$$

Information on the location of amplitude poles in the complex momenta planes, which correspond to zeroes of Jost function are then used to obtain 2 complex (or 4 real) equations:

$$D(k_1^r, k_2^r) = 0, \qquad D(k_1^R, k_2^R) = 0. \tag{16}$$

The superscripts $r$ and $R$ correspond to resonances $a_0(980)$ and $a_0(1450)$ which are included in the amplitude parameterization. Eqs.(16) along with information on $a_0(980)$ and $a_0(1450)$ branching fractions to $\pi\eta$ and $K\overline{K}$ channels are used to fix 5 parameters of the FSI amplitude [5].

## 4 Mass distributions and phase dependence

We define the double differential cross section as

$$\frac{d\sigma}{dtdM_{\pi\eta}} = \frac{1}{4}\frac{1}{(2\pi)^3}\frac{|k|}{32m^2 E_\gamma^2}\sum_{\lambda_\gamma,\lambda,\lambda'}|\langle\lambda'|A_{\pi\eta}|\lambda_\gamma\lambda\rangle|^2, \qquad (17)$$

where $m$ and $E_\gamma$ are the proton mass and photon energy in the laboratory frame respectively. For the calculation of the mass distribution we integrate Eq.(17) over the physical region of $t$. The calculation was performed at the photon energy $E_\gamma=7$ GeV which roughly corresponds to energies accessible in new and upgraded JLab experiments. As we are interested in the high energy regime we use the reggeised version of the photoproduction amplitude defined in Eq.(1). The striking feature of mass distribution in Fig. 3 is that interference of the real Born amplitude and the final state interaction terms results in the resonance being represented by the minimum (double dot-dash line). To develop a maximum, the background term needs to acquire the imaginary part. Similar feature was observed for the $a_0(980)$ production in $\gamma\gamma$ fusion [6]. Dot-dashed, dashed and dotted curves in Fig. 3 represent mass distributions resulting from interference of the final state interactions terms with the background amplitude "corrected" by additional phase of $\frac{\pi}{2}$, $\pi$ and $\frac{3}{2}\pi$ respectively. Detailed study of the size and origin of the imaginary part of the background will be likely possible when high statistics data on $\pi\eta$ photoproduction from CLAS12 and GlueX experiments will accessible.

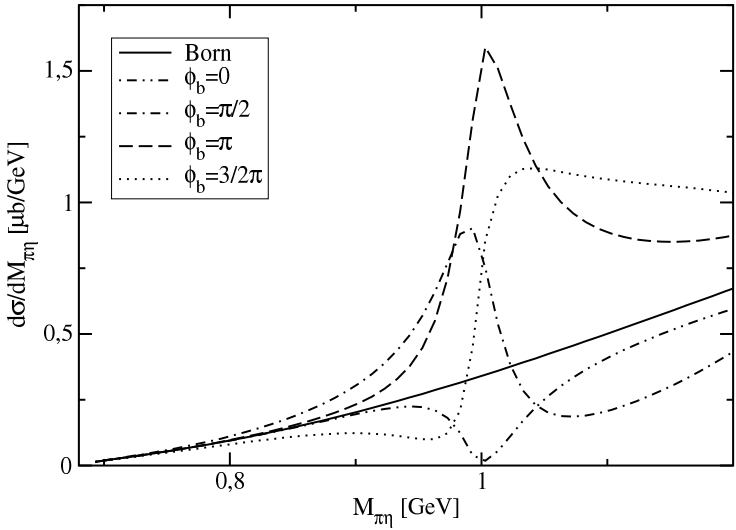

**Figure 3.** Dependence of the mass distribution on the phase of the background amplitude

## 5 Summary

We calculated the $M_{\pi\eta}$ mass spectrum in the region corresponding to isovector resonance $a_0(980)$ at energy of 7 GeV which will be accessible in upgraded JLab facilities. We have analyzed the interference pattern of resonance and background amplitude and found that for purely real Born background the $a_0(980)$ resonance is represented by the minimum in the mass distribution. The maximum can be developed only when the background acquires the imaginary part whose size can be estimated by experiment.

## Acknowledgments

This research has been funded by the Polish National Science Center (NCN) grant No. DEC-2013/09/B/ST2/04382